\documentclass{article}

\usepackage[final]{neurips_2021} 

\usepackage[utf8]{inputenc} 
\usepackage[T1]{fontenc}    
\usepackage{hyperref}       
\usepackage{url}            
\usepackage{booktabs}       
\usepackage{amsfonts}       
\usepackage{nicefrac}       
\usepackage{microtype}      
\usepackage{xcolor}         
\usepackage{graphicx}       

\title{CLP: A framework for dataset integration and exploitation}
\title{Homogenization of Existing Inertial-Based Datasets to Support Human Activity Recognition}

\author{%
  Hamza Amrani, Daniela Micucci, Marco Mobilio, Paolo Napoletano\\
  Department of Informatics, Systems and Communication\\
  University of Milano - Bicocca, Milan, ITALY\\
  \{\texttt{hamza.amrani, daniela.micucci, marco.mobilio, paolo.napoletano\}}\texttt{@unimib.it} 
}

\begin{document}

\maketitle


\begin{abstract}
Several techniques have been proposed to address the problem of recognizing activities of daily living from signals.
Deep learning techniques applied to inertial signals have proven to be effective, achieving significant classification accuracy. Recently, research in human activity recognition (HAR) models has been almost totally model-centric. 

It has been proven that the number of training samples and their quality are critical for obtaining deep learning models that both perform well independently of their architecture, and that are more robust to intraclass variability and interclass similarity. 
Unfortunately, publicly available datasets do not always contain hight quality data and a sufficiently large and diverse number of samples (e.g., number of subjects, type of activity performed, and duration of trials) Furthermore, 

%
datasets are heterogeneous among them and therefore cannot be trivially combined to obtain a larger set.

The final aim of our work is the definition and implementation of a platform that integrates datasets of inertial signals in order to make available to the scientific community large datasets of homogeneous signals, enriched, when possible, with context information (e.g., characteristics of the subjects and device position). The main focus of our platform is to emphasise data quality, which is essential for training efficient models.
\end{abstract}

\section{Introduction}
\label{section:introduction}

Human Activity Recognition (HAR) aims at automatically classifying activities performed by humans (including falls) by analyzing signals acquired by sensors~\cite{reyes2016transition,micucci2017falls, medrano2014detecting}.
%
Recent methods and approaches mostly exploit inertial sensors embedded in smartphones, smartwatches, fitness trackers, and ad-hoc wearable devices. 

In recent years, deep learning techniques have been successfully applied for 1D signals, exploiting their capability to overcome most of the issues raised by traditional machine learning techniques, thanks to their properties of local dependency and scale invariance~\cite{zeng2014convolutional}. While deep learning methods are powerful and achieve high performance, they rely on very complex models that depend on estimating a large number of parameters, which  in turn requires a considerable amount of available data~\cite{bianco2018benchmark}, whose quality increases the performance of the classification process.

Building an effective dataset is a complex task. Several factors undermine its goodness. These include the naturalness with which users perform the tasks, the position of the device, the balance of subjects involved, the number of samples recorded, and so on. Even if the dataset design is done in a rigorous way, there are factors that unfortunately are not controllable:   
%
%
the intraclass variability and the interclass similarity. The former means that different people perform the same activity in different ways, so a bijective association between signal and activity performed does not exist; the latter means that fundamentally different classes show very similar characteristics in terms of sensor data \cite{medrano2014detecting,lane2011enabling,krupitzer2018hips}.

In the literature, a number of HAR datasets are available~\cite{casilari2017analysis,8480322,8858508}. However these datasets, besides suffering from the above problems, are heterogeneous with each other. For example, signals are sometimes expressed in different units of measurement, may have different acquisition frequencies, and accelerations may include gravity or not. Moreover, ADLs and therefore the labels in the datasets, do not have a common dictionary or ontology, which leads in having similar labels that actually have different meanings. 
Thus, datasets cannot be used together without a significant effort to harmonise them. 

The availability of a dataset containing a large number of samples, also obtained from the integration of existing datasets, is a well-known issue both in the field of ADLs recognition from inertial sensors and in other domains, such as that related to image processing~\cite{ratner2017snorkel}.
In the context of ADLs recognition from inertial signals, Bartlett et al. proposed labels aggregation at the semantic level~\cite{bartlett2017acctionnet}. Labels of six existent datasets, resampled to 200 Hz using linear interpolation, were relabelled manually to reflect their semantic similarity, obtaining 13 different activity labels. The proposal however does not seem to consider details such as units of measurement or the presence of gravity in accelerations. Furthermore, the proposal allows only one configuration at 200 Hz. Recently, Siirtola et al. proposed a Matlab tool called Open HAR~\cite{siirtola2018openhar} that aggregates labels at a syntactic level but fails to consider the semantics of the original signals. Obinikpo et al. proposed a system for big data-d-health integration~\cite{obinikpo2019big}. They split the integration into different layers: data acquisition, data processing, analytics, and application. Nevertheless, the proposal is too general concerning the homogenization of different data sources since its major concern is handling missing values while integrating databases.

The main contribution of this paper is the definition of a homogenization procedure that allows to integrate heterogeneous datasets in order to obtain a larger dataset to be used for the definition of recognition techniques.
The procedure has been implemented and integrated in a platform termed Continuous Learning Platform (CLP) that
%
makes available (i) a large amount of labelled inertial signals related to ADLs and falls; (ii) a catalogue of downloadable activity recognition models, and (iii) a service that, given a set of raw data, identifies the corresponding ADL. The platform is available at the following URL:  \url{https://gitlab.com/Pervasive-Healthcare/CLP}.

The paper is organised as follows: Section~\ref{section:homogenization} describes the homogenization procedure we identfied; Section~\ref{section:implementation} provides an overview of the CLP platform and some implementation details; while Section 4 presents final remarks.

\section{Homogenization Procedure}
\label{section:homogenization}

Homogenization involves two different phases: \emph{signals homogenization} and \emph{labels homogenization}.

\textbf{Signals Homogenization}. The signals homogenization procedure, as the name suggests, focuses on signals and handles three types of inconsistencies: differences in sampling frequencies, discrepancies in units of measurement, and the presence of gravity for acceleration signals. The process involves three steps:
\begin{itemize}
	\item \textit{Frequency uniformation}. A resampling operation takes place intending to modify the frequency of the time series. There are two different types of resampling: upsampling, when the sampling rate is increased compared to the original one, and downsampling, when the sampling rate is decreased compared to the original one. The number of samples obtained with the new frequency can be calculated starting from Equation \ref{eq:numSamples}.
	\begin{equation}
		numSamples=\frac{numOriginalSamples * newFreq}{originalFreq}
		\label{eq:numSamples}
	\end{equation}
	Once the new time series with the desired frequency has been obtained, the respective timestamp is calculated for each sample starting from 0. 
	Information regarding the original sampling rate of the dataset is obtained by the metadata if available, or, if this information is not provided, it is calculated using the timestamp associated with the inertial data. 
	When the sampling frequency has been obtained or estimated, the data can be resampled.
	For all type of sensors, we adopted a frequency of 50Hz. Literature suggests that about
50Hz is a suitable sampling rate that permits to model human activities~\cite{50HZ}.
	
	\item \textit{Unit of measurement uniformation}. The unit of measurement, unlike the sampling frequency, must always be provided. It is not possible to trace this information from inertial data. The unit of measurement of a given sensor is converted to the desired one by a specific formula. For example, to convert an accelerometer from $g$ to $m/{s^2}$, it is necessary to add the gravitational acceleration, that is, multiplying by the gravitational acceleration constant, equal to $9.80665 m/{s^2}$.
	For accelerometer, gyroscope, and orientation we adopted $m/s^{2}$, $rad/s$, and $microTesla$ respectively. 
		
	\item \textit{Gravity uniformation}. To remove gravity and reduce noise or artefacts, a Butterworth filter is commonly applied~\cite{robertson2013research,takeda2014drift}. We considered a fifth-order 0.5 Hz low pass Butterworth filter with a Nyquist frequency of 25Hz, assuming the gravity force to have only low-frequency components.
	Since the information about gravity being included in the signal or not is often omitted, we decided to apply the Butterworth filter to all raw inertial signals.
\end{itemize}


\textbf{Labels Homogenization}. The label homogenization procedure aligns the labels of the ADLs in the dataset to be homogenized with the labels chosen as the reference ones.
%
%
Because there is no shared definition for each ADLs and each dataset can have different and conflicting labels, relying on the syntactic similarity may not be sufficient (e.g., walk and walking describe the same activity although the two labels are syntactically different, on the other hand, sitting, and sitting down, may be syntactically similar, but they refer to different actions: being sit and actively sitting from a standing position).
We provide two different approaches to make an accurate mapping:
\begin{itemize}
	\item \textit{Label Syntax Similarity (LSS)}.
	This index indicates the syntactic similarity of two strings (labels) without considering the signal component. We used the Levenshtein distance, which is the minimum number of elementary modifications (deleting a character, replacing one character with another, or inserting a character) that allows transforming a string A into another string B. For example, the Levenshtein distance between walk and walking equals 3.
	This distance can be helpful in some cases, but it cannot provide adequate information to automatically choose the correct mapping.
	
	\item \textit{Label Signal Similarity Distance (LSSD)}.
	Considering the signal component from the time series, we extracted 21 features from the magnitude component of the signal for each window. These features describe the different properties of the signal both in the time and frequency domain. The average of the features extracted from the windows of a specific activity represents the entire ADL. Each dataset has a feature vector per ADL.
	The Euclidean distance is then applied to determine possible associations. 
	If the distance of two activities is low, it suggests the associated signals are very similar, and the suggestion is a mapping between the two ADLs.
	
	\item \textit{Magnitude comparison graph}.
    The procedure also includes a manual comparison of the magnitude of different labels' time series for each pair that satisfies the minimum equality criteria. A random time series is taken for each label to create the graph. This is a visual aid to help a user in deciding which mappings to perform.
\end{itemize}

The final decision about mappings is always left to the users. However, LSS and the LSSD are useful Decision Support Systems that ease the process and reduce the confusion that may arise from heterogenous labels

\section{Continuous Learning Platform}
\label{section:implementation}

Continuous Learning Platform (CLP) is a freely available platform that implements the homogenization procedure described in Section~\ref{section:homogenization} and provides the tools to both integrate new datasets and to retrive homogenized datasets and recognition models.

Three main components constitutes CLP: \textit{i) Data Collection}, that acquires a new dataset; \textit{ii) Data Management} that homogenizates the new dataset and insert it in the incrementally built dataset; and \textit{iii) Data Distribution}, which enables users and applications to query the platform and obtain homogeneous sets of labelled signals, but also ad-hoc trained classifiers.

Two more components complete the platform: the \textit{Repository Manager} that deals with the management and internal storage of all the platform data, and a \textit{Web Application} responsible for making all the services offered by the platform available through an intuitive graphical interface. 

The individual components were developed in Python, while a message broker (RabbitMQ) has been used to permit efficient interaction between the components. The Web Application is built using the Angular framework, while REST APIs have been used for communication.
The components are implemented as a Flask server to manage the REST calls coming from the Web Application. The platform uses MongoDB as a database.

The rest of the section will give a more detailed view of the three main components.

%
\textbf{Data Collection} includes the following modules. The \textit{Dataset Loader} allows users to physically upload datasets and store them in a local repository.
The \textit{Driver Loader} handles the upload of datasets drivers, which are scripts provided by the users allowing the platform to correctly interpreter the data in the dataset. The \textit{Importer} executes the actual import by running the uploaded driver over the uploaded dataset, thus standardizing it according to a specific structure shared by all the datasets imported. It is noteworthy that, at this stage, the content of the datasets are still heterogeneous, while they share a common structure. 

%
%
\textbf{Data Management} reifies the homogenization procedure described in Section~\ref{section:homogenization}. It includes the following modules. The \textit{Data aligner} homogenizes each signal in a dataset at a specific frequency and unit of measurement, according to a common configuration. It also applies a Butterworth filter to remove the gravity from accelerations and to reduce noise and other artefacts. The \textit{Feature extractor} computes the magnitude and a set of hand-crafted features~\cite{ferrari2019hand} from inertial signals to determine signal similarity. Finally, the \textit{Label comparator} uniforms the dataset's labels to include a standard unified set, also considering the signal similarities. This module is semi-automatic: it provides suggestions on the assignment of labels, but ultimately it is up to the end-user to decide whether or not to accept the suggestions.

%
%
\textbf{Data Distribution} consists of the following modules. The \textit{Query builder} handles users' queries for homogenized data. The \textit{Classifier builder} trains classifiers according to user queries. Thanks to the homogenization procedure, the models generalize well in terms of inter and intra-subject variability regardless of the fact that are trained on data coming from different datasets. The \textit{Classifier deployer} distributes trained classifiers according to users' requests. The \textit{Online classifier} provides online services related to classification: given a set of inertial signals, it provides information regarding the subject's activity.

%
%

\section{Conclusions}
\label{section:conclusion}

Human Activity Recognition is a challenging and active research field and has seen rapid growth in the past few years.
The lack of large datasets reduces exploitation of deep learning techniques as they usually require large amounts of data that exceeds the size of individual datasets.

We propose a platform (CLP) that enables the integration and the distribution of data coming from heterogeneous sources.
The main components of CLP have been fully implemented and are available at  \url{https://gitlab.com/Pervasive-Healthcare/CLP}, while some aspects, such as the Web application is still under active development.

Eight datasets available in the literature have been already integrated and used to train a convolutional neural network and a recurrent neural network. These preliminary tests confirm the improvement of performances by combining existing datasets from different smartphones and various contexts and will be the focus of future work.

Future directions include polishing the Web application and an intensive test of the overall platform with real-world applications. Relying on our previous work on personalization~\cite{9412140}, we are currently working on designing a software component that interfaces the CLP platform and an Android application to develop personalized models. 

\bibliographystyle{unsrtnat}
\bibliography{bibliography}

\end{document}